\documentclass[9pt,twocolumn,twoside]{opticajnl}
\journal{opticajournal} 
\providecommand{\JournalTitle}[1]{#1}

\dates{}
\doi{}

\usepackage{xcolor}

\usepackage{xcolor}

\title{Generating intense attosecond pulses and vectorizing polarization states from laser-plasma interactions}

\author[1,2]{Panfei Geng}
\author[1,2,*]{Yipeng Wu}
\author[1,2]{Zhixin Fan}
\author[1,2]{Min Chen}
\author[1,2]{Longqing Yi}
\author[1,2]{Xiaohui Yuan}
\author[1,2]{Zhengming Sheng}
\author[3]{Warren B. Mori}
\author[3]{Chan Joshi}
\author[1,2]{Jie Zhang}

\affil[1]{State Key Laboratory of Dark Matter Physics, Key Laboratory for Laser Plasmas (MOE), Tsung-Dao Lee Institute \& School of Physics and Astronomy, Shanghai Jiao Tong University, Shanghai 201210, China}
\affil[2]{Collaborative Innovation Center of IFSA, Shanghai Jiao Tong University, Shanghai 200240, China}
\affil[3]{University of California Los Angeles, Los Angeles, CA 90095, USA}

\affil[*]{yipeng.wu@sjtu.edu.cn}

\begin{abstract}

Vector beams with spatially structured polarization and intertwined spin-orbital angular momentum (SAM-OAM) provide powerful degrees of freedom for tailoring light-matter interactions. While such structured beams are well established in the visible and infrared regimes, extending them to the extreme-ultraviolet (EUV) and soft X-ray (SXR) domains at relativistic intensities remains a major challenge. Here, we investigate the generation of higher-order harmonic vector beams driven by relativistic laser-plasma interactions. Combining theoretical analysis with three-dimensional particle-in-cell simulations, we elucidate the underlying physical mechanisms governing the transfer and conversion of polarization and orbital angular momentum during harmonic generation. We demonstrate that both the polarization topology and OAM of the emitted harmonics can be deterministically controlled by the topological charges of the driving field. Owing to the intrinsic properties of vector beams, either few-cycle driving pulses or vector polarization gating applied to multi-cycle pulses enable the production of intense isolated attosecond pulses featuring spiral wavefronts and spatially tailored polarization states. These results establish a pathway toward high-intensity structured light sources in the EUV and SXR regimes and open new opportunities for ultrafast and strong-field light-matter interaction studies with engineered angular momentum.

\end{abstract}

\setboolean{displaycopyright}{false} 

\begin{document}

\maketitle

\section{Introduction}



Angular momentum (AM), as an intrinsic property of light, comprises orbital angular momentum (OAM) and spin angular momentum (SAM). OAM is associated with the phase structure of light \cite{Allen_PRA_1992}, while SAM depends on the polarization states of light \cite{Poynting_1909}. Recently, with advancements in optical field manipulation, structured light \cite{Forbes_NP_2021,He_LSA_2022} has garnered significant interest, which refers to light tailored in all its degrees of freedom, especially phase and polarization. The most well-known examples of structured light are the vortex and vector beams. Vortex beam, exemplified by Laguerre-Gaussian (LG) modes carrying well-defined OAM \cite{Shen_Light_Sci_Appl_2019,Allen_PRA_1992}, exhibits helical phase structures and phase singularities. Vector beam, also referred to as the Poincaré beam, is characterized by spatially varying polarization states and polarization singularities \cite{Khajavi_JOP_2016, Guzman_JO_2018, Chen_SB_2018}. The simplest examples of the latter include radially and azimuthally polarized cylindrical vector beams, where the electric field is oriented either radially or azimuthally at all angular positions. Unlike vortex beams, which primarily possess OAM, vector beams exhibit complex OAM-SAM coupling. In general, vector beams can be viewed as a superposition of orthogonal circularly polarized LG vortex beams with different topological charges \cite{Milione_PRL_2011, Milione_PRL_2012, Guzman_JO_2018}. Both vortex and vector beams can be described by higher-order Poincaré spheres \cite{Forbes_NP_2021, Milione_PRL_2011, Milione_PRL_2012, Guzman_JO_2018}.

Owing to their uniqueness, vortex and vector beams have found important applications in many fields \cite{Shen_Light_Sci_Appl_2019, Shi_SCPMA_2024, Chen_SB_2018, Guzman_JO_2018}. For example, they are utilized in optical communication \cite{Wang_NP_2012, Wang_PR_2016, Ndagano_JLT_2018, Wang_COL_2017}, optical manipulation \cite{Grier2003, Tweezers__np_2011}, quantum entanglement \cite{Entanglement_2001, TwistedPhotons_2007}, and advanced imaging \cite{Maurer_LPR_2010, Furhapter_OL_2005, Furhapter_OE_2005, Chen_OL_2013} in the low-intensity domain. At relativistic intensities, the applications of such structured laser beams have also been extensively explored, including generation of strong magnetic field structures \cite{Longman_PRR_2021}, laser-driven acceleration of particles \cite{PayeurAPL2012, Vieira_PRL_2014, Zaim_PRL_2017, Wen_OE_2019, Yin_Shi_PRL_2021, wangwenpengPRL2020, zaimPRX2020, PowellPRL2024, Liberman_Nat_Commun2025}, suppression of laser-plasma instabilities \cite{Guo_MRE_2023}, generation of OAM-carrying and/or spin-polarized lepton beams \cite{BaumannaPhysPlasmas2018,VieiraPhysRevLett2018, suntingPRL_2024}.
To enable such diverse applications, numerous techniques for generating vortex and vector laser pulses have been proposed or demonstrated over the past two decades, particularly in high-intensity/-power regimes, such as mode conversion \cite{Shi_PRL_2014,Denoeud2017,wangwenpengPRL2020,WenpengWangCommunPhys2025}, polarization transformation \cite{zaimPRX2020}, spin-orbit conversion \cite{Qukenan2017}, chirped pulse amplification \cite{ChenZ_HPLSE_2022, Feng_US_2023}, optical parametric amplification \cite{zhongOptica2021}, and stimulated Raman and Brillouin amplification \cite{Vieira_NC_2016, Wu_NC_2024}. For example, high-power vortex beams with peak powers on the order of tens to hundreds of terawatts have been experimentally generated through spiral phase plate- or mirror-based mode conversion techniques  
\cite{Denoeud2017,wangwenpengPRL2020,LongmanOptLett2020,WenpengWangCommunPhys2025}. Similarly, radially and azimuthally polarized vector beams with comparable power scales have also been realized experimentally via polarization conversion using phase masks \cite{zaimPRX2020}. Upon focusing, such vortex and vector beams have achieved relativistic intensities ranging from $10^{19}$ to $10^{20}$ W/cm$^2$ \cite{Denoeud2017,wangwenpengPRL2020,zaimPRX2020,LongmanOptLett2020,WenpengWangCommunPhys2025}. However, the wavelengths of the resulting vortex and vector pulses are mainly in the visible and infrared spectral range. Extending such structured pulses into EUV and/or soft X-ray (SXR) range can not only expand the capability of existing applications, but also further open up completely new research directions in a wide range of areas, especially when combined with high beam intensities/powers. 

High harmonic generation (HHG) provides a particularly promising route for generating structured EUV and SXR radiation. In gases driven by near-infrared (near-IR) lasers, HHG can produce coherent EUV/SXR pulses\cite{CorkumPhysRevLett1993,LewensteinPhysRevA1994,KrauszRMP2009,PopmintchevScience2012} and  polarization-gating techniques can further confine the emission into isolated attosecond bursts \cite{TcherbakoffPhysRevA2003,KovacevEPJD2003,SolaNatPhys2006,MartensPhysRevA2004,SansoneScience2006,ChangPhysRevA2007}. Moreover, driving HHG with vortex or vector beams can yield short-wavelength radiation with spiral phase fronts and/or spatially varying polarizations  \cite{albadelasheras_optica_2022,CarlosPRL2013,GariepyPRL2014,GeneauxNC2016,Rego_Science_2019}. However, in gaseous media, the driving laser intensity must remain relatively low (typically $\lesssim 10^{14}\text{W/cm}^2$) to mitigate strong ionization, which in turn limits the achievable harmonic yield. To overcome this limitation, plasma-based HHG from solid-density targets driven by relativistic-intensity laser pulses has been proposed for generating high-output structured EUV/SXR pulses. Significant theoretical and experimental efforts have been devoted to generating such short-wavelength vortex beams via this mechanism  \cite{Denoeud2017,Wang_NC_2019, Li_NJP_2020, Yi_PRL_2021,Trines_NC_2024, Zhang_PRL_2015, Zhang_PRAppl_2021, Duff_SR_2020, Jirka_PRR_2021, Bacon_MRE_2022}. In comparison, research on vector-beam HHG in the relativistic regime remains much more limited. This disparity arises because vector beams are inherently more complex due to the coupling between OAM and SAM, as well as their intrinsically spatially varying polarization states, rendering their investigation particularly difficult and challenging. Although a limited number of studies have attempted to address this, they primarily focus on the simplest case of radially or azimuthally polarized vector beam with net (total) zero OAM \cite{Chen_PRA_2021}. To our best knowledge, general characteristics of high-intensity harmonic vector beams generated from laser-solid interactions, as well as fundamental mechanisms governing the conversion of AM and the evolution of polarization patterns, have not been systematically investigated, which are crucial for their diverse applications.

In this work, we investigate the generation of complex EUV and SXR vector harmonics by employing an intense near-infrared vector beam irradiating a flat overdense plasma. Through theoretical analysis and fully three-dimensional (3D) particle-in-cell (PIC) simulations, we provide the first comprehensive explanation of fundamental mechanisms during this process. We demonstrate that the OAM and polarization patterns of harmonic vector beams can be precisely controlled by adjusting the topological charge of incident laser pulses. Notably, for high-order vector harmonics carrying nonzero net OAM, both the intensity distribution and the electric field vectors exhibit longitudinal rotation. In addition, we show that by using a near- or few-cycle driving vector beam,
a high-intensity/-power spiral isolated attosecond pulse with a well-defined specific polarization pattern can be obtained. Furthermore, we propose a novel "vector polarization gating" technique, which also allows the generation of isolated attosecond vector pulse even though the incident laser beam has a relatively long duration. Our findings open new avenues for manipulating light's OAM and polarization degrees of freedom at EUV and SXR wavelengths, thereby enabling novel realms of light-matter interactions.

\section{Theoretical analyses}
\subsection{Characteristics analysis of fundamental vector beams}
To illustrate the underlying physics involved, we first conduct theoretical analyses. In general, a vector beam propagating along the $x$-direction can be viewed as a superposition of two LG modes with orthogonal circular polarization states and different topological charges \cite{Khajavi_JOP_2016,Guzman_JO_2018, Chen_SB_2018}:
\begin{equation}
\mathbf{E}_V=E_{LG,l_L}e^{i\Delta\varphi}\mathbf{e}_L+E_{LG,l_R}\mathbf{e}_R,
\end{equation} 
where $\mathbf{e}_L=\mathbf{e}_y-i\mathbf{e}_z$ and $\mathbf{e}_R=\mathbf{e}_y+i\mathbf{e}_z$ are the circular unitary vectors with left and right handedness, respectively. $E_{LG,l_L}$ or $E_{LG,l_R}$ is the electric field of the left- or right-handed circularly polarized LG mode with topological charge of $l_L$ or $l_R$, given by 
\begin{equation}
E_{LG,l}=E_{0}(\frac{\sqrt2 \rho }{w_0})^{|l|}\exp(-\frac{\rho ^2}{{w_0}^2})\sin^2(\frac{\pi\xi}{c\tau})\cos(\omega_{0}t-k_{0}x+l\phi),
\end{equation} 
with $l = l_L$ or $l_R$. Here $E_0$, $\omega_0$, $k_0 = \omega_0 / c$, $\phi = \arctan(z/y)$, $\rho = \sqrt{y^2 + z^2}$, $\xi = ct - x$, $w_0$ and $\tau$ denote the electric field amplitude, laser frequency, wave number, azimuthal angle, radial coordinate, comoving coordinate, spot size, and the full duration of the electric field envelope, respectively. Accordingly, the full width at half maximum (FWHM) of the intensity envelope is $0.36\tau$. Note that the wavefront curvature is ignored due to the interaction distance is much less than the Rayleigh length, and the zero radial mode index is assumed here.  $\Delta \varphi = \varphi_L-\varphi_R$ is the relative phase difference between $E_{LG,l_L}$ and $E_{LG,l_R}$. Substituting Eq. (2) into Eq. (1), the electric field of a vector beam can be expressed by
\begin{equation}
    \mathbf{E}_V \approx E_{V0}\sin^2(\frac{\pi \xi }{c\tau })\sin(\omega_{0}t-k_{0}x+\frac{l_L+l_R}{2}\phi+\frac{\Delta\varphi+\pi}{2})\mathbf{e}_V,
\end{equation}
where $E_{V0} \varpropto  \exp(-\rho ^2/{w_0}^2)$ is the electric field amplitude and $\mathbf{e}_V$ denotes the electric field vector, which is given by
\begin{equation}
    \mathbf{e}_V=\cos(\frac{l_L-l_R}{2}\phi+\frac{\Delta\varphi}{2})\mathbf{e}_y+\sin(\frac{l_L-l_R}{2}\phi+\frac{\Delta\varphi}{2})\mathbf{e}_z.
\end{equation}
Note that Eq. (3) is derived in the near field by neglecting the difference in the radial amplitudes of LG modes with different topological charges. This approximation is valid when the radial intensity maxima of the two components sufficiently overlap. For equal beam waists ($w_{0L} = w_{0R} = w_0$), this requires $|l_L| \approx |l_R|$. More generally, the overlap is maximized when $w_{0R} \approx w_{0L} \sqrt{|l_L| / |l_R|}$. When this condition is not satisfied, the reduced overlap leads to a lower conversion efficiency (see Section A of Supplement 1). Equations (3)-(4) express the phase and complex polarization of the vector beam in a separable form, which, as will be seen subsequently,
greatly simplifies the generation and characteristics analysis of harmonic vector beams. From Eqs.(3)-(4), we can clearly see that there are two key parameters that determine the characteristics of the vector beam, $l_{V}=(l_L+l_R)/2$ and $l_C=(l_L-l_R)/2$. The parameter $l_V$ is the topological charge of the fundamental vector beam, also known as the topological Pancharatnam charge \cite{Niv_OE_2006}. This charge is directly associated with the total OAM or net OAM carried by the vector beam. The parameter $l_C$ determines the polarization pattern of the vector beam \cite{Khajavi_JOP_2016}. Specifically, the pattern is star (or web) for $l_C<0$, while lemon (or flower) for $l_C>0$.  Meanwhile, except for $l_C=+1$, the number of radial lines $N$ in the polarization pattern (i.e., the number of angular sectors partitioned by disclination lines) is also determined by the parameter $l_C$ as $N=|2(l_C-1)|$\cite{Khajavi_JOP_2016}.

\subsection{Generation and characteristics analysis of harmonic vector beams}
\begin{figure}[ht]
\centering
\fbox{\includegraphics[width=0.66\linewidth]{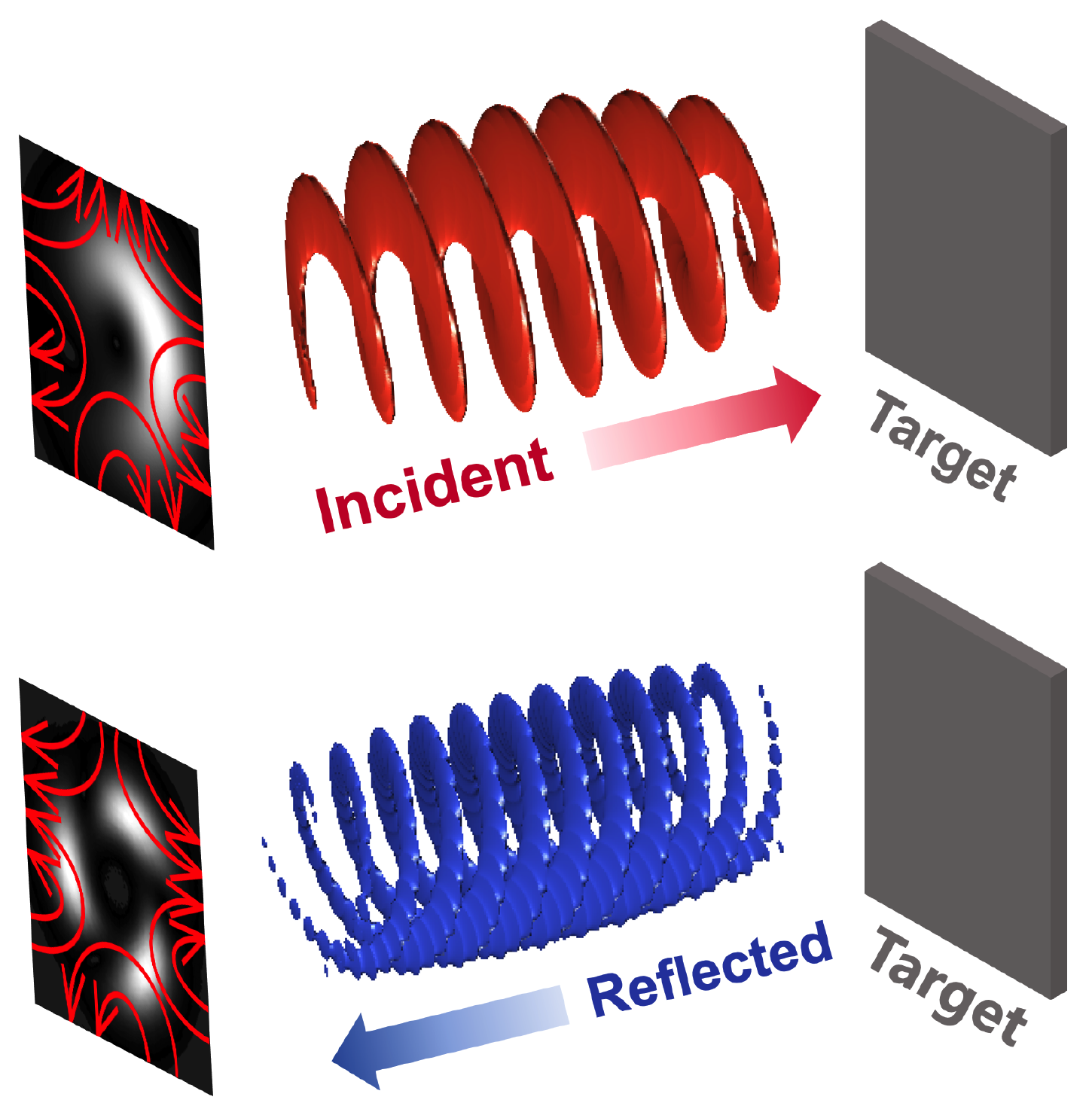}}
\caption{Schematic view: Generation of intense harmonic vector beams from laser-solid interactions. When an intense near-infrared vector pump beam (red) is normally incident on an overdense planar target, the resulting longitudinal ponderomotive force drives relativistic oscillations of electrons at the critical density surface. These oscillations reflect the incident light and emit high-order harmonic vector radiation (dark blue). The slices represent the intensity and polarization patterns of the incident and reflected harmonic vector beams in the transverse plane.}
\label{fig_diagram}
\end{figure}

HHG from laser-solid interaction is an effective way to produce intense short-wavelength radiation. When a relativistic linearly polarized laser pulse irradiates the target, it induces strong oscillations of the critical surface. This oscillating surface acts as a mirror to reflect the incident light and emit linearly-polarized high-order harmonics, which has been elucidated clearly with the well-known relativistic oscillating mirror (ROM) model \cite{Lichters_pop_1996,Bulanov_pop_1994,Baeva_PRE_2006,Thaury_NP_2007}. However, in our case, the driving laser is a SAM-OAM-coupled vectorially polarized beam (described by Eqs. (3)-(4)). Consequently, the conventional ROM transforms into a "relativistic vector oscillating mirror" (RVOM), which inherits the SAM-OAM information encoded in the driving beam. For simplicity, we consider the case of a vector beam normally incident on a flat target, as illustrated in Fig. \ref{fig_diagram}. In this configuration, if the laser spot size is much bigger than the laser wavelength, the RVOM dynamics are primarily governed by the radiation pressure (longitudinal ponderomotive force) of the vector beam, 
\begin{equation}
F_{px} \sim \frac{\partial I_V}{\partial_x} \sim \frac{\partial |\mathbf{E}_V|^2}{\partial_x} \sim \sin(2\omega_{0}t+2l_V\phi-2k_0x+\Delta\varphi+\pi),
\label{Equation_5}
\end{equation}
where $I_V \propto |\mathbf{E}_V|^2$ is the laser intensity. When neglecting other forces such as the space charge force from the ion background and only considering $F_{px}$ as the driving force, the electron layer (critical surface) of RVOM will undergo periodic oscillations, with the oscillation displacement $X$ estimated as
\begin{equation}
    X(t)\sim X_0\sin(2\omega_{0}t+2l_V\phi+\Delta\varphi+\pi),
    \label{Equation_6}
\end{equation}
where $X_0$ represents the oscillation amplitude. Note that the exact analytical solution for $X$ is very difficult to obtain. Here, we only adopt a first-order approximation, which is sufficient to analyze the main characteristics of the harmonics. Equation (6) indicates that the critical surface oscillates longitudinally at a frequency of $2\omega_0$, and the oscillating phase is related to the azimuthal angle $\phi$ when $l_V \neq 0$. This oscillating surface layer then reflects the incident vector pulse. The reflected field observed at ($x_{obs},t$) is emitted at a retarded time $t_{ret}$ from RVOM located at $X(t_{ret})$. Here, the retarded time $t_{ret}$ is defined by $t_{ret}=t-X(t_{ret})/c+x_{obs}/c$. Therefore, according to Eq. (3), the electric field at the observer can be written (strictly) as
\begin{equation}
\mathbf{E}_V(x_{obs},t) \sim \sin \left(\omega_0 t + l_V\phi - k_0 X(t_{ret}) + \frac{\Delta\varphi+\pi}{2}\right)\mathbf{e}_V.
\label{Equation_7}
\end{equation}
Here, for simplicity, we set the observation point at $x=x_{obs}=0\mu$m. Strictly speaking, the $X(t_{ret})$ term in Eq. (7) should be taken from Eq. (6) as $X(t_{ret})\sim X_0\sin(2\omega_{0}t-2k_0X(t_{ret})+2l_V\phi+\Delta\varphi+\pi)$, which is implicit and requires an iterative solution. To obtain an analytical expression for the radiation field, we adopt a first-order approximation $X(t_{ret})\approx X(t) \sim X_0\sin(2\omega_{0}t+2l_V\phi+\Delta\varphi+\pi)$. Substituting this expression into Eq. (7) and employing the Jacobi-Anger identity \cite{Cuyt_Springer_2008}, we obtain
\begin{equation}
    \mathbf{E}_V \sim \sum_{n = 0}^{\infty} J_n(\epsilon) \sin\left[(2n+1)\left(\omega_0 t + l_V \phi + \frac{\Delta\varphi + \pi}{2}\right)\right] \mathbf{e}_V,
\end{equation}
where $J_n$ denotes the Bessel function of the first kind and $\epsilon=-k_0X_0$. Note that here we made the approximation $J_{2n-1}(\epsilon)-J_{2n}(\epsilon)\approx J_{2n-1}(\epsilon )$ and $J_{2n}(\epsilon)+J_{2n+1}(\epsilon)\approx J_{2n}(\epsilon)$, since $\epsilon\ll 1$. From Eq. (8), we can see that the reflected field contains only odd harmonics, which is similar to the case of normally incident linearly polarized light \cite{Lichters_pop_1996}. Meanwhile, we can clearly see that the OAM carried by the $q$-th harmonic is $ql_V\hbar=q(l_L+l_R)\hbar/2$. Note that from the perspective of photons, we can draw the same conclusion based on the OAM conservation. For the $q$-th harmonic, we have $q=n_R+n_L$, where $n_L(n_R)$ is the number of left(right)-handed circularly polarized photons absorbed by the harmonic. Because of symmetry, $q$ must be an odd integer, i.e., $n_R-n_L=\pm 1$. Therefore, we have ($n_{R}=(q+1)/2$, $n_L=(q-1)/2$) for right circular polarization (RCP) channel, and ($n_{R}=(q-1)/2$, $n_L=(q+1)/2$) for left circular polarization (LCP) channel. Therefore, for the $q$-th harmonic, we have $l_{q,R}=(q+1)l_R/2+(q-1)l_L/2$ in RCP channel, and $l_{q,L}=(q-1)l_R/2+(q+1)l_L/2$ in LCP channel. Thus, the total topological charge of the $q$-th harmonic is $l_q=(l_{q,R}+l_{q,L})/2=q(l_L+l_R)/2=ql_V$. Moreover, as implied by Eq. (8), the polarization of the harmonic vector beam is the same as that of the fundamental vector beam, which is determined by $\mathbf{e}_V$ given by Eq. (4). Therefore, both OAM and polarization patterns of harmonic vector beams can be conveniently controlled by tuning the parameters of the incident beam.

Note that the characteristics of the vector beam are significantly different from those of circularly or linearly polarized LG beams. For vector beams, both the intensity $I_V$ ($I_V \propto |\mathbf{E}_V|^2 \sim [1- \cos (2\omega_0t-2k_0x+2l_V\phi+\Delta\varphi+\pi)]$) and 
ponderomotive force $F_{px}$ (see Eq. (5)) show time-dependent high-frequency oscillations and azimuthal dependence. In contrast, circularly polarized LG beams do not possess these characteristics although they also exhibit spatially-varying polarizations. On the other hand, although linearly polarized LG beams share similarities with vector beams in that their intensity and ponderomotive force contain time-dependent high-frequency oscillations and are azimuthally dependent, their polarization patterns are spatially uniform. Therefore, from this particular perspective, vector beams can be seen as a synthesis of the characteristics inherent in both circularly and linearly polarized LG beams.

\section{Simulation results}
\subsection{Characteristics of harmonic vector beams}
To confirm our theoretical analysis as well as to visualize the characteristics of the harmonic vector beam, three-dimensional (3D) particle-in-cell (PIC) simulations are performed with the OSIRIS code \cite{Fonseca_OSIRIS_2002}. The size of the simulation box is 15$\mu$m($x$)$\times$24$\mu$m($y$)$\times$24$\mu$m($z$) with $1875\times480\times480$ grid cells. The time step is $dt=0.017$ fs. The driving vector beam has a wavelength of $\lambda_0=2\pi c/\omega_0 =0.8 \mu$m and a peak electric field of $E_{V0,peak}=2.2\times10^{13}$ V/m, which corresponds to a peak normalized vector potential of $a_0=eE_{V0,peak}/m_e\omega_0c=5.5$. Its electric-field envelope follows a $\sin^2(\pi t/\tau)$ profile over the interval [0,$\tau$], where $\tau = 40$ fs is the full duration, corresponding to a FWHM duration (intensity) of $0.36\tau = 14.40$ fs. The focal spot size is $w_0=4\mu $m. Four sets of typical topological charges ($l_L$, $l_R$) are selected and the corresponding vector beam characteristics are summarized in Tab. \ref{tab1}. The plasma target consists of a slab with a thickness of $2\lambda_0$ and an exponentially decaying (with distance) pre-plasma, $n_e(x)=n_0$exp$[(x-x_0)/L]$. Here, $x_0=13.0\mu$m, $n_0=100n_c$ and $L=0.2\lambda_0$, with $n_c = 1.744\times 10^{21}$cm$^{-3}$ being the critical density for the driving laser.
Furthermore, in order to improve the quality of the harmonic vector beams as well as to maintain their polarization patterns, we employ a pre-plasma truncation at a cutoff density of $n_{cut}=20n_c\approx 4a_0n_c$ \cite{B_Y_Li_PRE_2019, B_Y_Li_PRL_2022} in simulations. Such truncations have been experimentally demonstrated via surface plasma compression \cite{B_Y_Li_PRL_2022}. 
Both electrons and C$^{6+}$ ions are mobile, and each grid contains 8 macroparticles. Absorbing boundary conditions are used for both fields and particles. The standard Yee solver is used.  

\begin{table*}[htbp]
\caption{Characterization of vector beams for different topological charges ($l_L$, $l_R$)}
\label{tab1}
\centering
\begin{tabular}{cccccc}
\hline
($l_L$, $l_R$) & $l_V$ & $l_C$ & Polarization pattern & N & OAM for the $q$-th harmonic \\
\hline
(-1, 1) & 0 & -1 & Web & 4 & 0 \\
(3, -3) & 0 & 3 & Flower & 4 & 0 \\
(-1, 2) & 1/2 & -3/2 & Web & 5 & $q\hbar/2$ \\
(2, 1) & 3/2 & 1/2 & Flower & 1 & $3q\hbar/2$ \\
\hline
\end{tabular}
\caption*{Four typical values of  ($l_L$, $l_R$) are selected: the first two have no net OAM ($l_V = 0$), while the other two have net OAM ($l_V\neq 0$), with each type of $l_V$ including $l_C < 0$ and $l_C > 0$. Note that OAM in the last column is for the laser electric field, and OAM should be doubled for the intensity. All these characterizations are confirmed by the 3D PIC simulation results in Fig. \ref{fig_vector_HHG_main}.}
\end{table*}

\begin{figure}[h]
\centering
\fbox{\includegraphics[width=\linewidth]{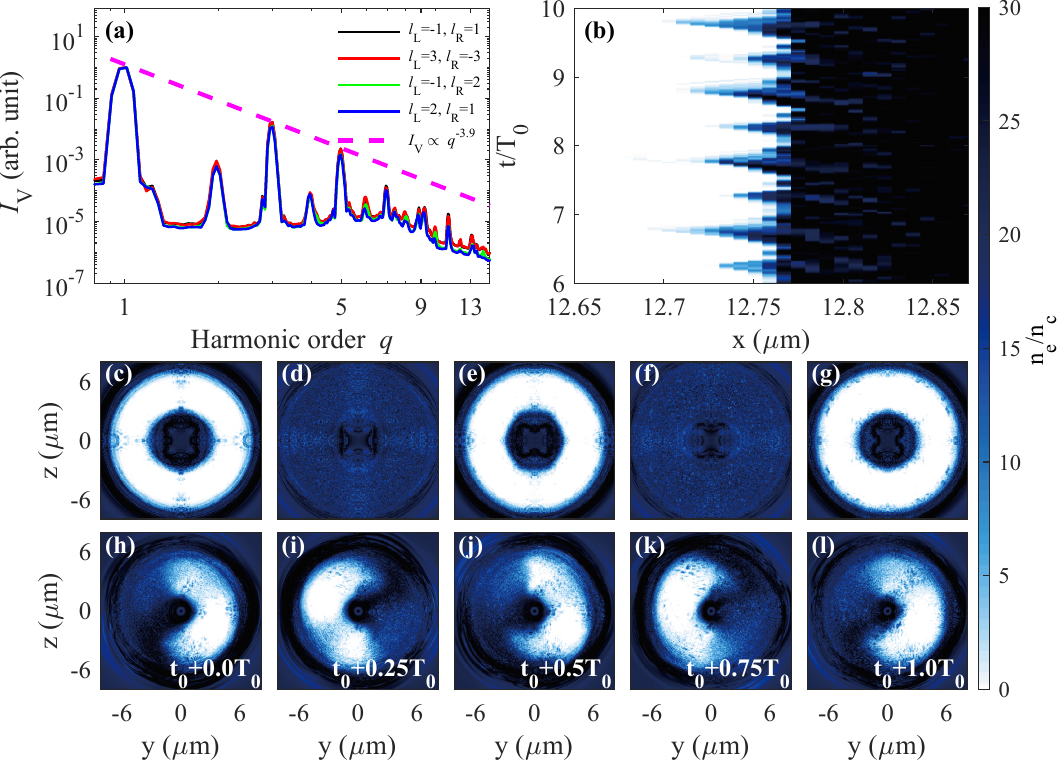}}
\caption{(a) Spectra of reflected pulses for different vector pump beams. The magenta dashed line shows the spectral scaling law. (b) Spatiotemporal evolution of electron density at a fixed transverse  position ($y=w_0, z=0$). (c)-(l)  Transverse distribution of electron density at $x=12.76\mu$m for different moments, with (c)-(g) corresponding to ($l_L=3, l_R=-3$) and (h)-(l) corresponding to ($l_L=-1, l_R=2$).}
\label{fig_spectrum_density}
\end{figure}

Figure \ref{fig_spectrum_density}(a) shows the spectrum of reflected pulses for four vector beams. For all cases, the spectrum mainly contains odd harmonics, which is consistent with the analysis based on RVOM model. The spectral scaling law $I_{V} \propto q^{-3.9}$, as indicated by the magenta dashed line, implies a relatively high energy conversion efficiency. Note that the weaker even harmonics and modulation of spectral envelope are caused by the pre-plasma truncation \cite{B_Y_Li_PRE_2019}. Under normal incidence on a planar target with a smooth pre-plasma profile, i.e., in the absence of pre-plasma truncation, even-order harmonics are strongly suppressed. As the laser intensity increases, intense harmonic vector beams with higher orders can be generated (see Section B of Supplement 1). Figure \ref{fig_spectrum_density}(b) shows the spatiotemporal evolution of electron density at a fixed transverse position ($y=w_0, z=0$). As one can see, the compact electron sheet mainly oscillates at twice the laser frequency ($2\omega_0$) and has a very short excursion distance ($\sim 0.05\mu$m) in the presence of pre-plasma truncation. For comparison, we show the harmonic spectra and spatiotemporal density evolution without pre-plasma truncation in Section C of Supplement 1. There, the even harmonics and modulation almost disappear, and the spectral width of each harmonic is significantly broader, due to a more diffuse electron sheet and a longer excursion distance ($\sim 0.2\mu$m). Note that the pre-plasma density scale length is known to significantly affect the conversion efficiency of high-order harmonics \cite{KahalyPRL2013,DollarPRL2013,JahnOptica2019}. For vector beams, however, this dependence is weak in the case of a truncated pre-plasma, as considered here. By contrast, untruncated exponential profiles show a much stronger scale-length sensitivity (see Section D of Supplement 1). Figures \ref{fig_spectrum_density}(c)-\ref{fig_spectrum_density}(g) show that the electron density is independent of the azimuthal angle $\phi$ for $l_V=0$ ($l_L=3, l_R=-3$), with only a time dependence of $\sim \sin (2\omega_0t)$. While Figs. \ref{fig_spectrum_density}(h)-\ref{fig_spectrum_density}(l) show that the density has a clear dependence on $\phi$ in addition to its temporal oscillation ($\sim \sin(2\omega_0t+2l_V\phi)$) when $l_V \neq 0$ ($l_L=-1, l_R=2$).  

\begin{figure}[ht]
\centering
\fbox{\includegraphics[width=\linewidth]{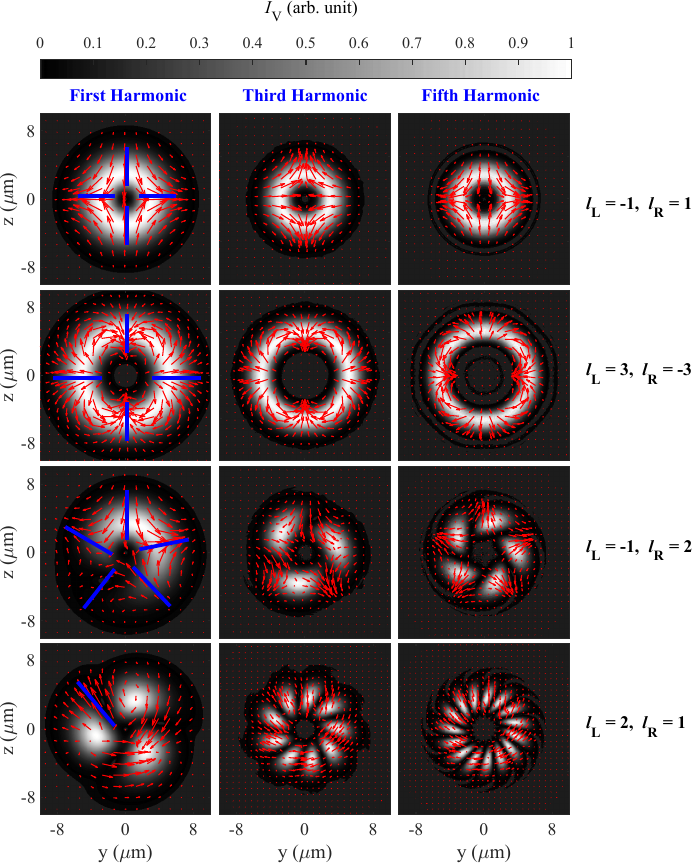}}
\caption{Characterization of harmonic vector beams extracted from 3D PIC simulations. The grayscale colormap shows the intensity distribution $I_V \propto E_{Vy}^2 + E_{Vz}^2$ of the vector beam in the transverse $(y,z)$ plane, while the red arrows represent the electric-field vectors. Blue solid lines mark radial lines of the polarization pattern. Different rows represent different types of the vector beam, corresponding to Tab. \ref{tab1}, while different columns are for different harmonic orders.}
\label{fig_vector_HHG_main}
\end{figure}

The harmonic characteristics of four typical vector beams are shown in Fig. \ref{fig_vector_HHG_main}. In order to clearly show the net OAM it carries, we present the beam intensity $I_V \propto |\mathbf{E}_V|^2 = E_{Vy}^2 + E_{Vz}^2$, encoded in grayscale, instead of the individual electric-field components ($E_{Vy}$ or $E_{Vz}$) in the transverse plane.
\begin{figure}[ht]
\centering
\fbox{\includegraphics[width=\linewidth]{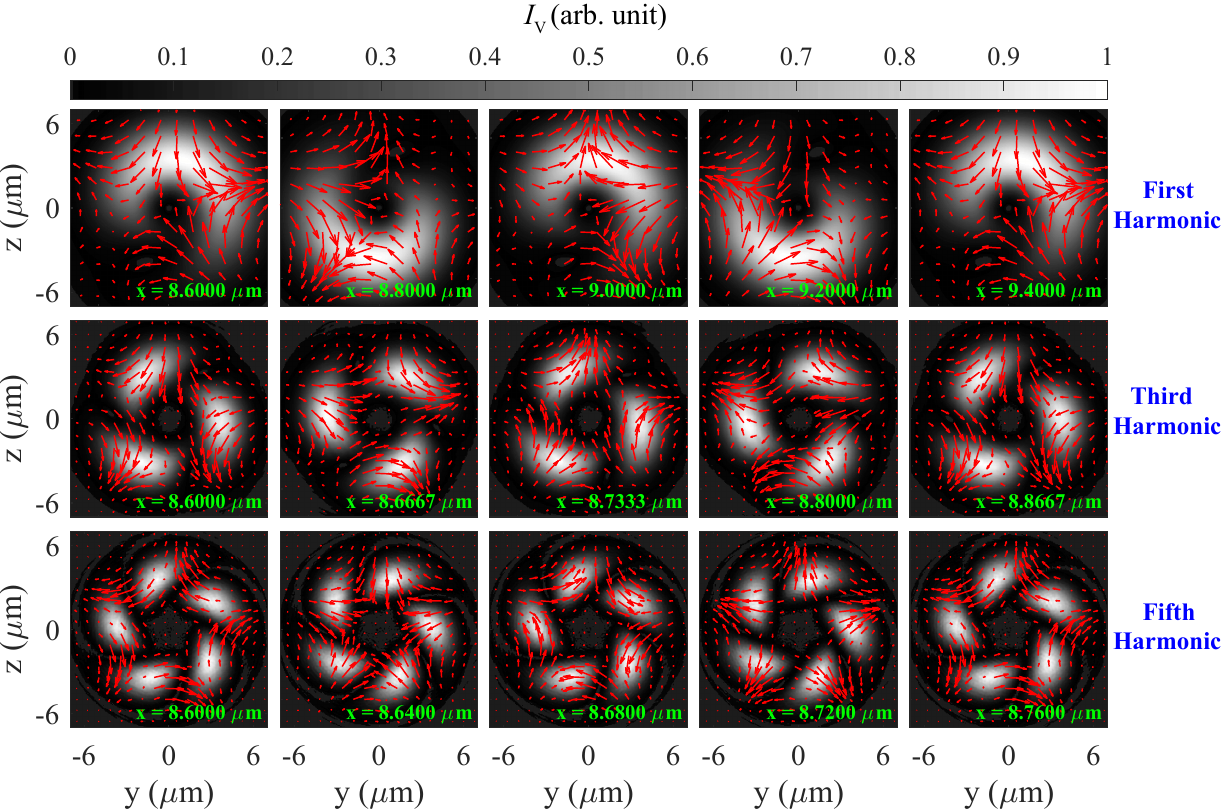}}
\caption{Intensity and local electric field vectors of harmonic vector beams rotate with the longitudinal position. The parameter ($l_L = -1$, $l_R = 2$) is used. The grayscale colormap shows the intensity distribution, while the red arrows represent the electric-field vectors. Different rows are for different harmonic orders. For the $q$-th harmonic, intensity and local electric field vectors rotate with periods $\lambda_0/2q$ and $\lambda_0/q$, respectively.}
\label{fig_pattern_rotation}
\end{figure}
The red arrows are the electric field vectors (polarization patterns) and blue solid lines mark the radial lines in patterns. As one can see, for ($l_L=-1$, $l_R=1$) and ($l_L=3$, $l_R=-3$), all harmonics have no net OAM, which is confirmed by the absence of lobes in the intensity pattern, and is consistent with the theory ($l_{V}=0$). However, their polarization patterns are clearly distinct. The pattern is 'Web' for the former due to $l_C = -1 < 0$, while 'Flower' for the latter due to $l_C = 3 > 0$. For both cases, as mentioned earlier, the number of radial lines $N$ in the pattern satisfies $N=|2(l_C-1)|=|l_L-l_R-2|$. The last two rows of Fig. \ref{fig_vector_HHG_main} indicate that for ($l_L=-1$, $l_R=2$) and ($l_L=2$, $l_R=1$), all harmonics carry the net OAM and are in good agreement with the theoretical analysis. For example, for the 5-th harmonic of ($l_L=-1$, $l_R=2$), the number of lobes (OAM) in the intensity is $2ql_V=q(l_L+l_R)=5$. Note that the OAM in the intensity is twice that in the electric field, as mentioned earlier. To further confirm these results, we directly calculate the SAM ($S_{x,y,z}$), OAM ($L_{x,y,z}$), and total angular momentum (TAM, $J_{x,y,z}$) values of harmonic vector beams using the electromagnetic fields extracted from the PIC simulations (see Section E of Supplement 1). The calculation results validate the existence of the longitudinal OAM and are consistent with the theoretical values ($L_x \approx J_x\approx ql_V\hbar  = q(l_L+l_R)\hbar /2$, $S_{x,y,z}\approx L_{y,z}\approx J_{y,z}\approx 0$, see Table S1 in Supplement 1). We have also performed a grid-resolution study, which verifies that the extracted SAM/OAM is numerically robust and approaches convergence with refined spatial and temporal resolution. Furthermore, the polarization patterns are well preserved (as mentioned earlier, it is determined by the parameter $l_C=(l_L-l_R)/2$). All these results confirm that both the OAM and the spatially varying polarization of the harmonic vector beam can be conveniently controlled by the incident beam ($l_L$, $l_R$). 

Furthermore, Eq. (8) implies that the harmonic intensity $I_V$ and local electric field vectors rotate with the longitudinal position (or time). As an example, we show the case of ($l_L = -1$, $l_R = 2$) in Fig. \ref{fig_pattern_rotation}. We can see that for the $q$-th harmonic, the rotational period of the intensity and local electric field vectors are $\lambda_0/2q$ and $\lambda_0/q$, respectively.
While the overall polarization pattern (i.e., 'Web' with 5 radial lines) remains unchanged due to $l_C=(l_L-l_R)/2=-3/2<0$. Note that vector beams are different from circularly or linearly polarized LG light and combine the characteristics of both, as mentioned in Section 2 (See Section F of Supplement 1). Such a rotation of intensity and local electric field vectors might be useful for dynamic probing of polarization-dependent systems or for generating monoenergetic electron sheets \cite{Yin_Shi_PRL_2021}.  

\subsection{Generation of an isolated attosecond vector pulse}\label{isolated}
As the cornerstone of attosecond science \cite{KrauszRMP2009}, isolated attosecond pulses are important tools for studying and controlling the ultrafast dynamical processes in matter. However, it is still challenging to generate such pulses, especially combined with high intensity and structured polarization. Our proposed scheme is also a promising route to generating intense isolated attosecond pulses with specific polarization when driven by a near-single or few-cycle vector beam.

\begin{figure}[h]
    \centering
    \includegraphics[width=\linewidth]{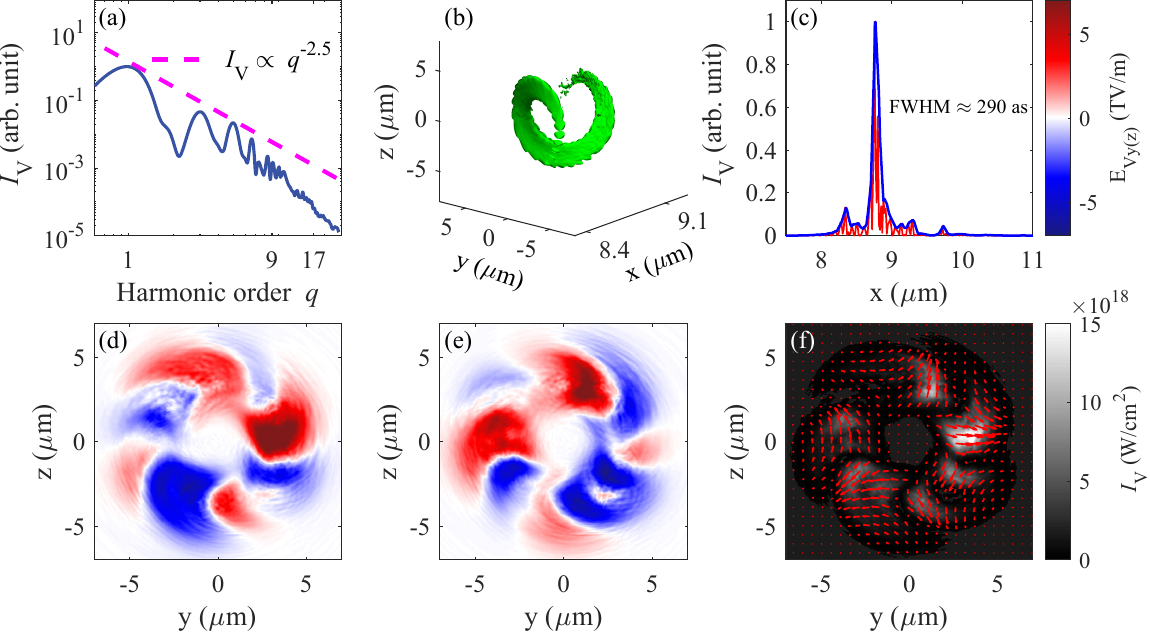}
    \caption{Generation of an isolated intense attosecond vector pulse driven by a near-single-cycle laser pulse with a FWHM duration of 2.40 fs. (a) Harmonic spectrum of the reflected light. The magenta dashed line indicates the spectral scaling law. (b) Intensity isosurface $(I_V/I_{V,max}=0.3)$ of the reflected vector beam filtered for the 5th-19th harmonics. (c) Lineout of the filtered field intensity at $y=-3.0 \mu$m and $z=2.16 \mu$m. (d-f) Transverse distributions at $x=9.0 \mu$m of the (d) $E_{Vy}$ and (e) $E_{Vz}$ electric-field components, and (f) the corresponding intensity $I_V \propto E_{Vy}^2 + E_{Vz}^2$, with in-plane electric-field vectors shown by red arrows.}
    \label{fig_few_cycle_3fs}
\end{figure}

For example, we consider a driver with $a_0 = 12$, a 2.40-fs (FWHM) duration, $l_L = -1$, and $l_R = 2$. Figure \ref{fig_few_cycle_3fs} summarizes the properties of the reflected vector pulse. Figure \ref{fig_few_cycle_3fs}(a) presents the harmonic spectrum of the reflected light, with the magenta dashed line denoting the spectral scaling law. By filtering the 5th-19th harmonics, we synthesize an attosecond pulse as shown in Fig. \ref{fig_few_cycle_3fs}(b-f). The intensity isosurface in Fig. \ref{fig_few_cycle_3fs}(b) and the longitudinal distribution in Fig. \ref{fig_few_cycle_3fs}(c) confirm the generation of an isolated 290 as pulse exhibiting a helical structure. Furthermore, the transverse slices in Figs. \ref{fig_few_cycle_3fs}(d-f) reveal that the pulse carries a net OAM and a spatially varying polarization (web-like pattern). Notably, reducing the driver amplitude to $a_0 = 5.5$ while extending its FWHM duration to 5.40 fs yields a single vector pulse of 870 as (see Section G of Supplement 1). Such isolated attosecond pulses with both spatially tailored polarization states and spiral wavefronts have the potential to revolutionize the attosecond science. They can play an important role in light-matter interaction, such as ultrafast imaging and kinetic probing in polarization-dependent systems.

However, the driving vector beams currently available from high-power laser systems are typically multicycle. Here, we propose a vector polarization gating (VPG) technique to generate an isolated attosecond pulse from a relatively long vector beam. As addressed above, the vector beam is a superposition of two orthogonal circularly polarized LG components. For normal incidence onto a flat target, a single component does not generate harmonics, whereas their vector superposition does; therefore, when a relative delay $\tau_d$ is introduced, harmonic emission is confined to their temporal overlap. 

\begin{figure}[h]
\centering
\fbox{\includegraphics[width=\linewidth]{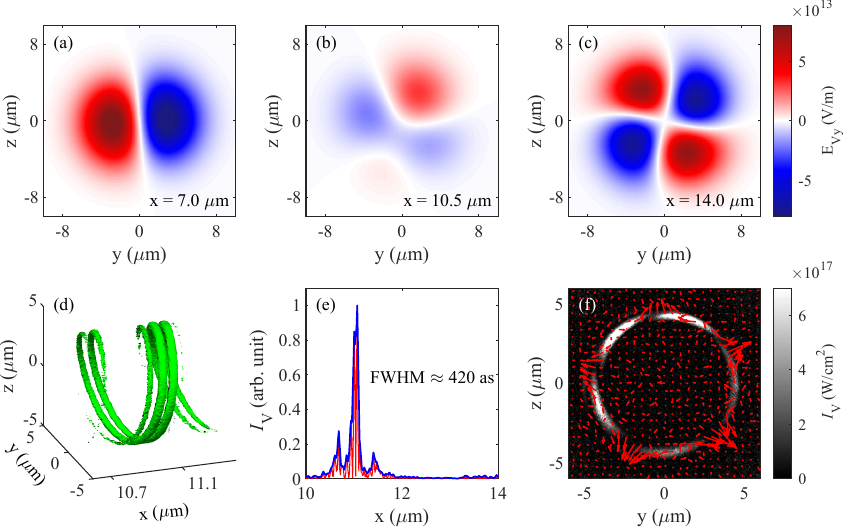}}
\caption{Generation of an isolated intense attosecond vector pulse using the polarization gating technique. (a)-(c) Transverse distribution of the electric field $E_{Vy}$ at different longitudinal positions for the incident pulse. (d) Intensity isosurface ($I_{V}/I_{V,max}=0.2$) of reflected pulse after filtering out harmonic components with $q \geq 5$. 
(e) The line plot of the intensity distribution for the filtered field. (f) Transverse distribution of the intensity for the filtered field at $x = 10.85 \mu m$. Red arrows are the electric field vectors. An isolated attosecond spiral pulse with a specific polarization pattern is generated by the polarization gating technique.}
\label{fig_polarization_gating}
\end{figure} 

Figure \ref{fig_polarization_gating} illustrates the simulation results of the VPG for multi-cycle incident beams. The topological charge of two incident LG beams is ($l_L=-1 $, $l_R=2$), with the intensity of $a_{0L}=a_{0R}=20$, full duration of $\tau=33$ fs and relative delay of $\tau_d=23$ fs. The density of the uniform plasma slab is 50 $n_c$. The top row of Fig. \ref{fig_polarization_gating} shows the transverse distribution of the electric field $E_{Vy}$ at different longitudinal positions for the incident pulse. As one can see, at the pulse tail (Fig. \ref{fig_polarization_gating}(a)) and head (Fig. \ref{fig_polarization_gating}(c)), where no overlap occurs, the topological charges are $l_L=-1$ and $l_R=2$, respectively. In contrast, at the pulse center (Fig. \ref{fig_polarization_gating}(b)), the overlapping of two LG beams generates a vector beam with a topological charge of $l_V=(l_L+l_R)/2=1/2$. The bottom row of Fig. \ref{fig_polarization_gating} shows the characteristics of the reflected pulse after filtering. Apparently, similar to the near-single-cycle incident pulse case (Fig. \ref{fig_few_cycle_3fs}), a high-intensity isolated attosecond pulse ($\approx $ 420 as) with a helical structure is obtained and the spatially varying polarization (web pattern) is maintained. The total energy of the incident pulse is 31.7 J, and the energy of the isolated attosecond vector pulse after spectral filtering is approximately 4.1 mJ, corresponding to an energy conversion efficiency of about $0.0129\%$. Details on the distinction between our VPG and conventional gating schemes \cite{TcherbakoffPhysRevA2003,KovacevEPJD2003,SolaNatPhys2006,MartensPhysRevA2004,SansoneScience2006,ChangPhysRevA2007}, together with an analytical estimate of the gating window, are provided in Section H of Supplement 1. Supplementary simulations further show that the VPG technique remains robust against small delay jitter (e.g., $\pm 3$ fs) and strong amplitude imbalance between the two driving components (e.g., $a_{0L}/a_{0R}=0.2$ for $a_{0R}=20$); see Section I of Supplement 1 for details.

\section{Conclusion and Discussion}\label{}
In summary, we have systematically investigated the characteristics of high-order vector harmonics generated from laser-solid interactions. Based on theoretical analysis and 3D PIC simulations, we demonstrate that the OAM and polarization states of harmonic vector beams can be precisely controlled by tuning the topological charge of incident laser pulses. By using a few-cycle incident vector beam, an isolated attosecond pulse with a spiral structure and spatially tailored polarization state can be generated. We have also proposed a novel "vector polarization gating" technique, which enables isolated attosecond vector pulse generation even though the incident beam has a relatively long duration with multi cycles. These structured pulses can bring new insights into light-matter interactions.

The high-intensity drive laser used in our scheme can be obtained by mode conversion \cite{Shi_PRL_2014,Denoeud2017,wangwenpengPRL2020,LongmanOptLett2020}, polarization transformation \cite{zaimPRX2020}, spin-orbit conversion \cite{Qukenan2017},  chirped pulse amplification \cite{ChenZ_HPLSE_2022,Feng_US_2023}, optical parametric amplification \cite{zhongOptica2021}, and plasma-based amplification \cite{Vieira_NC_2016, Wu_NC_2024}, as mentioned in the Introduction part. For example, vector lasers with peak intensities of up to $4.8\times 10^{19}$ W/cm$^2$ (normalized vector potential $a_0 \approx 4.7$) have already been experimentally generated through polarization transformation using phase masks, which can convert the laser polarization from linear to radial or azimuthal \cite{zaimPRX2020}. Therefore, this method has high experimental feasibility for current high-power laser systems. This feasibility is further supported by recent progress in ultrahigh-power sub-two-cycle laser drivers \cite{VeiszNatPhoton2025} and in advanced HHG beamlines and diagnostics relevant to relativistic regimes \cite{ShirozhanUltrafastSci2024}. Meanwhile, we note that the proposed scheme is analyzed primarily under an idealized configuration: a vector beam formed by two co-propagating LG components with equal amplitudes and a fixed relative phase is normally incident on a flat solid target. In realistic experiments, however, such ideal conditions cannot be maintained perfectly. To examine the robustness of the scheme against these imperfections, we carried out supplementary studies on small-angle oblique incidence, residual even-order harmonic leakage, plasma surface denting, amplitude mismatch, phase jitter, and small-angle beam crossing (see Sections J-M of Supplement 1). A preliminary experimental setup is proposed in Section N of Supplement 1. Furthermore, a reflective flat target is used in our scenario for simplicity. For a transmissive configuration (such as a foil with a small aperture \cite{Yi_PRL_2021}) or the tailored structured targets \cite{Trines_NC_2024}, there may be new degrees of freedom to control the characteristics of harmonic vector beams. In addition, similar to Refs. \cite{Rego_Science_2019, Wang_PRA_2022}, harmonic vector beams with time-varying OAM and/or polarization states may also be worth studying.

\begin{backmatter}
\bmsection{Funding}
This work was supported by the National Natural Science Foundation of China (Grant No. 12595365, 12475156, 12475246, 12225505, 12442501, and 12135009), the Fundamental and Interdisciplinary Disciplines Breakthrough Plan of Ministry of Education of China (Grant No. JYB2025XDXM204), and the Science and Technology Commission of Shanghai Municipality (Grant No. 25DX1400100 and 25JD1403300). The UCLA colleagues (Warren B. Mori and Chan Joshi) were supported by UCLA and the United States Department of Energy (Grant No. DE-SC0010064-0011). 

\bmsection{Acknowledgment} 
The simulations in this paper were performed on the $\pi$ 2.0 cluster supported by the Center for High Performance Computing at Shanghai Jiao Tong University.

\bmsection{Disclosures} 
The authors declare no conflict of interest.

\bmsection{Data Availability Statement} 
The data that support the findings of this study are available from the corresponding author upon reasonable request.

\bmsection{Supplemental document}
See Supplement 1 for supporting content.
\end{backmatter}

\bibliography{reference_main}



\ifthenelse{\equal{\journalref}{aop}}{%
\section*{Author Biographies}
\begingroup
\setlength\intextsep{0pt}
\begin{minipage}[t][6.3cm][t]{1.0\textwidth} 


  \noindent
  {\bfseries John Smith} received his BSc (Mathematics) in 2000 from The University of Maryland. His research interests include lasers and optics.
\end{minipage}
\begin{minipage}{1.0\textwidth}


  \noindent
  {\bfseries Alice Smith} also received her BSc (Mathematics) in 2000 from The University of Maryland. Her research interests also include lasers and optics.
\end{minipage}
\endgroup
}{}

\end{document}